\def\year{2022}\relax
\documentclass[letterpaper]{article} 
\usepackage{aaai22}  
\usepackage{times}  
\usepackage{helvet}  
\usepackage{courier}  
\usepackage[hyphens]{url}  
\usepackage{graphicx} 
\usepackage{natbib}  
\usepackage{caption} 
\DeclareCaptionStyle{ruled}{labelfont=normalfont,labelsep=colon,strut=off} 
\usepackage{algorithm}
\usepackage{algorithmic}
\usepackage{subcaption}
\usepackage{graphicx}
\usepackage{booktabs}
\usepackage{multirow}
\usepackage{newfloat}
\usepackage{listings}
\floatstyle{ruled}
\newfloat{listing}{tb}{lst}{}
\floatname{listing}{Listing}
\usepackage{bibentry}
\nocopyright 

\title{PDAugment: Data Augmentation by Pitch and Duration Adjustments for Automatic Lyrics Transcription}  
\author{
    Chen Zhang\textsuperscript{\rm 1},
    Jiaxing Yu\textsuperscript{\rm 1},
    LuChin Chang\textsuperscript{\rm 1},
    Xu Tan\textsuperscript{\rm 2},
    Jiawei Chen\textsuperscript{\rm 3},
    Tao Qin\textsuperscript{\rm 2},
    Kejun Zhang\textsuperscript{\rm 1}
    
}
\affiliations{

    \textsuperscript{\rm 1}Zhejiang University, China \\
    \textsuperscript{\rm 2}Microsoft Research Asia \\
    \textsuperscript{\rm 3}South China University of Technology \\
    zc99@zju.edu.cn, yujxzju@gmail.com, changluchin@gmail.com, xuta@microsoft.com,\\
    csjiaweichen@mail.scut.edu.cn, taoqin@microsoft.com, zhangkejun@zju.edu.cn
}

\begin{document}

\maketitle

\begin{abstract}
Automatic lyrics transcription (ALT), which can be regarded as automatic speech recognition (ASR) on singing voice, is an interesting and practical topic in academia and industry. ALT has not been well developed mainly due to the dearth of paired singing voice and lyrics datasets for model training. Considering that there is a large amount of ASR training data, a straightforward method is to leverage ASR data to enhance ALT training. However, the improvement is marginal when training the ALT system directly with ASR data, because of the gap between the singing voice and standard speech data which is rooted in music-specific acoustic characteristics in singing voice. In this paper, we propose PDAugment, a data augmentation method that adjusts pitch and duration of speech at syllable level under the guidance of music scores to help ALT training. Specifically, we adjust the pitch and duration of each syllable in natural speech to those of the corresponding note extracted from music scores, so as to narrow the gap between natural speech and singing voice. Experiments on DSing30 and Dali corpus show that the ALT system equipped with our PDAugment outperforms previous state-of-the-art systems by 5.9\% and 18.1\% WERs respectively, demonstrating the effectiveness of PDAugment for ALT.
\end{abstract}

\emph{ \textbf{Content Areas} --\ automatic lyrics transcription; data augmentation; automatic speech recognition; singing voice. }\rm

\section{Introduction}\label{sec:introduction}
Automatic lyrics transcription (ALT), which recognizes lyrics from singing voice, is useful in many applications, such as lyrics-to-music alignment, query-by-singing, karaoke performance evaluation, keyword spotting, and so on. ALT on singing voice can be regarded as a counterpart of automatic speech recognition (ASR) on natural speech. Although ASR has witnessed rapid progress and brought convenience to people in daily life in recent years~\citep{graves2006connectionist,graves2012sequence,chan2016listen,park2019specaugment,li2020towards,xu2020lrspeech}, there is not an ALT system that has the same level of high accuracy and robustness as the current ASR systems.
The main challenge of developing a robust ALT system is the scarcity of available paired singing voice and lyrics datasets that can be used for the ALT model training. To make matters worse, compared with ASR, ALT is a more challenging task -- the same content accompanied with different melodies will produce different pitches and duration, which leads to the sparsity of training data and further aggravates the problem of the lack of data. 
Though a straightforward method is to use speech data to enhance the training data of ALT, the performance gain is not large because there are significant differences between speech and singing voice. For example, the singing voice have some music-specific acoustic characteristics \citep{kruspe2016bootstrapping,mesaros2009adaptation} (details can be seen in Section~\ref{sec:speech_vs_singingvoice}) -- the large variation of syllable duration and highly flexible pitch contours are very common in singing, but rarely be seen in speech~\citep{tsai2018transcribing}. 

Previous work has already made some attempts in using speech data to improve the performance of ALT system:
~\citet{fujihara2006automatic} pretrain a recognizer with speech data and then built a language model containing the vowel sequences in the lyrics to further introduce the knowledge from the music domain. However, it only took advantage of the semantic information from lyrics and did not consider the acoustic properties of singing voice. 
~\citet{mesaros2009adaptation} adapted a pre-trained GMM-HMM based speech recognizer to singing voice domain by speaker adaptation technique, but it only shifted the means and variances of GMM components based on global statistics without considering the local information, resulting in very limited improvement. 
Some other work~\citep{kruspe2015training,basak2021end} tried to artificially generate “song-like” data from speech for model training.
\citet{kruspe2015training} applied time stretching and pitch shifting to natural speech in a random manner, which enriches the distribution of pitch and duration in “songified” speech data to a certain extent. Nonetheless, the adjustments are random, so there is still a gap between the patterns of “songified” speech data and those of real singing voice.
Compared to \citet{kruspe2015training}, \citet{basak2021end} further took advantage of real singing voice data. It transferred natural speech to singing voice domain with the guidance of real opera data. But it only took the pitch contours into account, ignoring duration, another key characteristic. Besides, it directly replaced the pitch contours with those of the real opera data, without considering the alignment of the note and syllable, which may result in the low quality of synthesized audio.

In this paper, we propose PDAugment, a syllable-level data augmentation method by adjusting pitch and duration under the guidance of music scores to generate more consistent training data for ALT training.
In order to narrow the gap between the adjusted speech and singing voice, we adjust the speech at syllable level to make it more in line with the characteristics of the singing voice. We try to make the speech more closely fit the music score, so as to achieve the effect of “singing out” the speech content.
PDAugment adjusts natural speech data by the following steps: 1) extracts note information from music scores to get the pitch and duration patterns of music; 2) aligns the speech and notes at syllable level; 3) adjusts the pitch and duration of syllables in natural speech to match those in aligned note. By doing so, PDAugment can add the information of music-specific acoustic characteristics into adjusted speech in a more reasonable manner, and narrow the gap between adjusted speech and singing voice.

Our contributions can be summarized as follows:
\begin{itemize}
    \item We develop PDAugment, a data augmentation method to enhance the training data for ALT training, by adjusting the pitch and duration of natural speech at syllable level with note information extracted from real music scores.
    \item We conduct experiments in two singing voice datasets: DSing30 dataset and Dali corpus. The adjusted LibriSpeech corpus is combined with the singing voice corpus for ALT training.
    In DSing30 dataset, the ALT system with PDAugment outperforms previous state-of-the-art system and Random Aug by 5.9\% and 7.8\% WERs respectively. In Dali corpus, PDAugment outperforms them by 24.9\% and 18.1\% WERs respectively. Compared to adding ASR data directly into the training data, our PDAugment has 10.7\% and 21.7\% WERs reduction in two datasets.
    \item We analyze the adjusted speech by statistics and visualization, and find that PDAugment can significantly compensate the gap between speech and singing voice. At the same time, the adjusted speech can keep the relatively good quality (the audio samples can be found in supplementary materials).
\end{itemize}

\section{Background}
In this section, we introduce the background of this work, including an overview of previous work related to automatic lyrics transcription and the differences between speech and singing voice.

\subsection{Automatic Lyrics Transcription}
The lyrics of a song provide the textual information of singing voice and are as important as the melody when contributing to the emotional perception for listeners~\citep{ali2006songs}. Automatic lyrics transcription (ALT) aims to recognize lyrics from singing voice.
In automatic music information retrieval and music analysis, lyrics transcription plays a role as important as melody extraction~\citep{hosoya2005lyrics}. 
However, ALT is a more challenging task than ASR -- not like speech, in singing voice, the same content aligned with different melodies will have different pitches and duration, which results in the sparsity of training data. So it is more difficult to build an ALT system than an ASR system without enough training data.

Some work took advantage of the characteristics of music itself:
\citet{gupta2018automatic} extended the length of pronounced vowels in output sequences by increasing the probability of a frame with the same phoneme after a certain vowel frame because there are a lot of long vowels in singing voice.
\citet{kruspe2016bootstrapping} boosted the ALT system by using the newly generated alignment~\citep{mesaros2008automatic} of singing and lyrics.
\citet{gupta2020automatic} tried to make use of the background music as extra information to improve the recognition accuracy.
However, they just designed some hard constraints or added extra information according to the knowledge from the music domain, and still did not solve the problem of data scarcity.

Considering the lack of singing voice database, some work aimed at providing a relatively large singing voice dataset:
\citet{dabike2019automatic} collected DSing dataset from real-world user information.
\citet{demirel2020automatic} built a cascade pipeline with convolutional time-delay neural networks with self-attention based on DSing30 dataset and provided a baseline for ALT task. 
Some other work leveraged natural speech data for ALT training: they based on pre-trained automatic speech recognition models and then made some adaptations to improve the performance on singing voice:
\citet{fujihara2006automatic} built a language model containing the vowel sequences in the lyrics but only used the semantic information from lyrics and ignored the acoustic properties of singing voice. \citet{mesaros2009adaptation} used speaker adaptation technique by shifting the GMM components only with global statistics but did not consider the local information.

However, singing voice has some music-specific acoustic characteristics which are not in speech, limiting the performance when training the ALT system directly with natural speech data. Some work tried to synthesize “song-like” data from natural speech to make up for this gap:
\citet{kruspe2015training} generated “songified” speech data by time stretching, pitch shifting, and adding vibrato. However, the degrees of these adjustments are randomly selected within a range, without using the patterns in real music.
\citet{basak2021end} took use of the F0 contours in real opera data and converted the speech to singing voice through style transfer. Specifically, they decomposed the F0 contours from the real opera data, obtained the spectral envelope and the aperiodic parameter from the natural speech, and then used these parameters to synthesize the singing voice version of the original speech. Nonetheless, in real singing voice, the note and the syllable are often aligned, but \citet{basak2021end} did not perform any alignment of the F0 contours of singing voice with the speech signal. This misalignment may lead to the change of pitch within a consonant phoneme (in normal circumstances, the pitch only changes between two phonemes or in vowels), which further causes distortion of the synthesized audio and limits the performance of ALT system.
Besides, they only adjusted the F0 contours, which is not enough to narrow the gap between speech and singing voice.

In this paper, we propose PDAugment, which improves the above adjustment methods by using real music scores and syllable-level alignment to adjust the pitch and duration of natural speech, so as to solve the problem of insufficient singing voice data.

\subsection{Speech vs. Singing Voice}
\label{sec:speech_vs_singingvoice}
In a sense, singing voice can be considered as a special form of speech, but there are still a lot of discrepancies between them~\citep{loscos1999low,mesaros2013singing,kruspe2014keyword}. These discrepancies make it inappropriate to transcribe the lyrics from singing voice using a speech recognition model trained on ASR data directly. In order to demonstrate the discrepancies, we randomly select 10K sentences from LibriSpeech~\citep{panayotov2015librispeech} for speech corpus and Dali~\citep{meseguer2019dali} for singing voice dataset, and make some statistics on them. 
The natural speech and singing voice mainly differ in the following aspects and the analysis results are list in Table~\ref{tab:statistics}.

\paragraph{Pitch} We extract the pitch contours from singing voice and speech, and compare the range and smoothness of the pitch. Here we use semitone as the unit of pitch. \\
\textit{\underline{Pitch Range}} Generally speaking, the range of the pitch in singing voice is larger than that in speech. \citet{loscos1999low} has pointed out that the frequency range in singing voice can be much larger compared to that in speech.
For each sentence, we calculate the pitch range (the maximum pitch value minus the minimum pitch value in this sentence). After averaging the pitch range of overall 10K sentences in the corpus, the average values are listed in \textit{Pitch Range} in Table~\ref{tab:statistics}. \\
\textit{\underline{Pitch Smoothness}}
The pitch of each frame in a certain syllable (when it is corresponding to a note) in singing voice remains almost constant whereas in speech the pitch changes freely along with the audio frames in a syllable. We call the characteristic of maintaining local stability within a syllable as \textit{Pitch Smoothness}. Specifically, we calculate the pitch difference between every two adjacent frames in a sentence and average it across the entire corpus of 10K sentences. The smaller the value of \textit{Pitch Smoothness}, the smoother the pitch contour. The results can be seen in \textit{Pitch Smoothness} of Table~\ref{tab:statistics}.

\paragraph{Duration} We also analyze and compare the range and stability of the syllable duration in singing voice and speech. The duration of each syllable varying a lot along with the melody in singing voice. While in speech, it depends on the pronunciation habits of the certain speaker. \\
\textit{\underline{Duration Range}} For each sentence, we calculate the difference between the duration of the longest syllable and shortest syllable as the duration range. The average values of the duration ranges in the entire corpus are shown as \textit{Duration Range} in Table~\ref{tab:statistics}. \\
\textit{\underline{Duration Variance}} We calculate the variance of the duration of syllables in each sentence and average the variances of all sentences in the whole corpus. The results are listed as \textit{Duration Variance} in Table~\ref{tab:statistics} to reflect the flexibility of duration in singing voice.

\begin{table}[!h]
\centering
\small 
\begin{tabular}{l | c c }
\toprule
Property & \textit{Speech} & \textit{Singing Voice} \\
\midrule
\textit{Pitch Range (semitone)} & 12.71 & 14.61  \\
\textit{Pitch Smoothness} & 0.93 & 0.84 \\
\textit{Duration Range (s)} & 0.44 & 2.40 \\
\textit{Duration Variance} & 0.01 & 0.11 \\
\bottomrule
\end{tabular}
\caption{The differences of acoustic properties between speech and singing voice.}
\label{tab:statistics}
\end{table}

Besides the differences in the characteristics we mentioned above, sometimes singers may add vibrato in some long vowels or make artistic modifications to the pronunciation of some words to make them sound more melodious, though it will result in a loss of intelligibility. Considering that some characteristics are hard to be quantified, in this work, we start with pitch and duration to build a prototype and propose PDAugment to augment ALT training data by adjusting pitch and duration of speech at syllable level according to music scores.

\section{Method}
In this section, we introduce the details of our proposed PDAugment: a data augmentation method by music score-guided syllable-level pitch and duration adjustments for automatic lyrics transcription. We first describe the overall pipeline of the ALT system and then introduce the designs of each component in PDAugment respectively.

\begin{figure}[thb]
    \centering
    \includegraphics[width=0.45\textwidth]{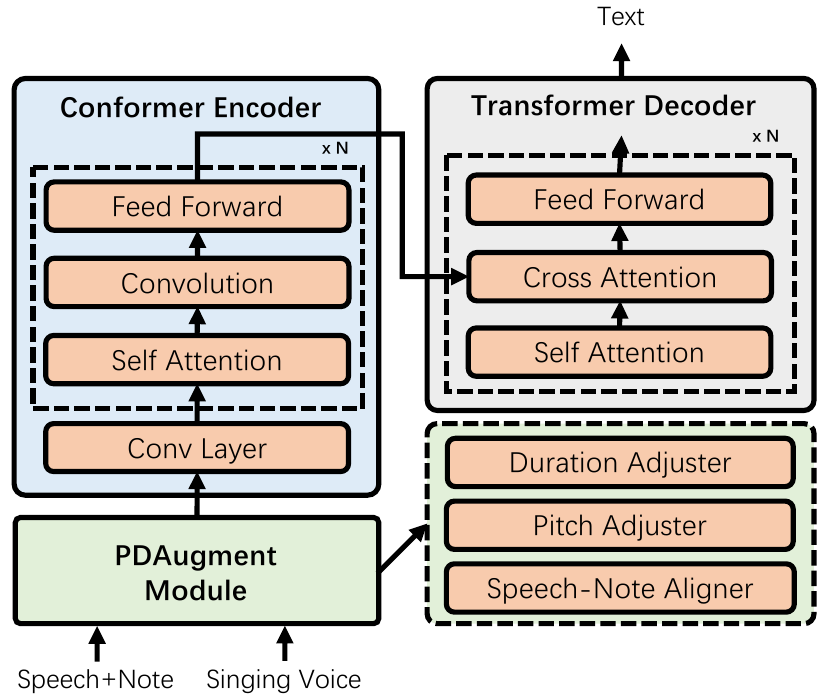}
    \caption{The overall pipeline of the ALT system equipped with PDAugment.}
    \label{fig:pipeline}
\end{figure}

\subsection{Pipeline Overview}
For automatic lyrics transcription, we follow the practice of existed automatic speech recognition system~\citep{watanabe2018espnet} and choose Conformer encoder~\citep{gulati2020conformer} and Transformer decoder~\citep{vaswani2017attention} as our basic model architecture. Different from standard ASR systems, we add a PDAugment module in front of the encoder as shown in Figure~\ref{fig:pipeline}, to apply syllable-level adjustments to pitch and duration of the input natural speech according to the information of aligned notes. When the input of ALT system is singing voice, we just do not enable PDAugment module. When the input is speech, PDAugment module takes the note information extracted from music scores as extra input to adjust the pitch and duration of speech, and then adds the adjusted speech into training data to enhance the ALT model.
The loss function of the ALT model consists of decoder loss $\mathcal{L}_{dec}$, and ctc loss (on top of encoder) $\mathcal{L}_{ctc}$~\citep{wang2020semantic}: 
\label{eq:loss}
$\mathcal{L} = (1-\lambda) \mathcal{L}_{dec} + \lambda \mathcal{L}_{ctc}$,
where $\lambda$ is a hyperparameter to trade-off the two loss terms. 
Considering that lyrics may contain more musical-specific expressions which are rarely seen in natural speech, the probability distributions of lyrics and standard text are quite different. We train the language model with in-domain lyrics data and then fuse it with ALT model in the beam search of decoding stage. 

We try to make the speech fit the patterns of singing voice more naturally as well as to achieve the effect of “singing out” the speech by PDAugment, so we adjust the pitch and duration of speech at syllable level according to those of corresponding notes in music scores instead of applying random adjustments.
To do so, we propose PDAugment module, which consists of three key components: 1) speech-note aligner, which generates the syllable-level alignment to decide what the corresponding note of a certain syllable is for subsequent adjusters; 2) pitch adjuster, which adjusts the pitch of each syllable in speech according to that of aligned notes; and 3) duration adjuster, which adjusts the duration of each syllable in speech to be in line with the duration of the corresponding notes. We introduce each part in the following subsections.

\subsection{Speech-Note Aligner}
According to linguistic and musical knowledge, in singing voice, syllable can be viewed as the smallest textual unit corresponds to note. 
PDAgument adjusts the pitch and duration of natural speech at syllable level under the guidance of note information obtained from the music scores. In order to apply the syllable-level adjustments, we propose speech-note aligner, which aims to align the speech and note (in melody) at syllable level. 

The textual content can serve as the bridge of aligning note of melody with the speech. Specifically, our speech-note aligner aligns the speech with note (in melody) in the following steps: \\
1) In order to obtain the syllable-level alignment of text and speech, we first convert the text to phoneme by an open-source tool\footnote{https://github.com/bootphon/phonemizer} and then align the text with speech audio by the Montreal forced aligner (MFA)~\citep{mcauliffe2017montreal} tool\footnote{https://github.com/MontrealCorpusTools/Montreal-Forced-Aligner} at phoneme level. Next, we group several phonemes into a syllable according to the linguistic rules~\citep{kearns2020does} and get the syllable-level alignment of text and speech. \\
2) For the syllable-to-note mappings, we set one syllable to correspond to one note by default, because in most cases of singing voice, one syllable is aligned with one note. Only when the time length ratio of the syllable in speech and the note in melody exceeds the predefined thresholds (we set the 0.5 as lower bound and 2 as upper bound in practice), we generate one-to-many or many-to-one mappings to prevent audio distortion after adjustments. \\
3) We aggregate the syllable-level alignment of text and speech, and the syllable-to-note mappings to generate the syllable-level alignment of speech and note (in melody) as the input of the pitch and duration adjusters.

\subsection{Pitch Adjuster}
\label{sec:aug_pitch}
\begin{figure}[thb]
    \centering
    \includegraphics[width=0.45\textwidth]{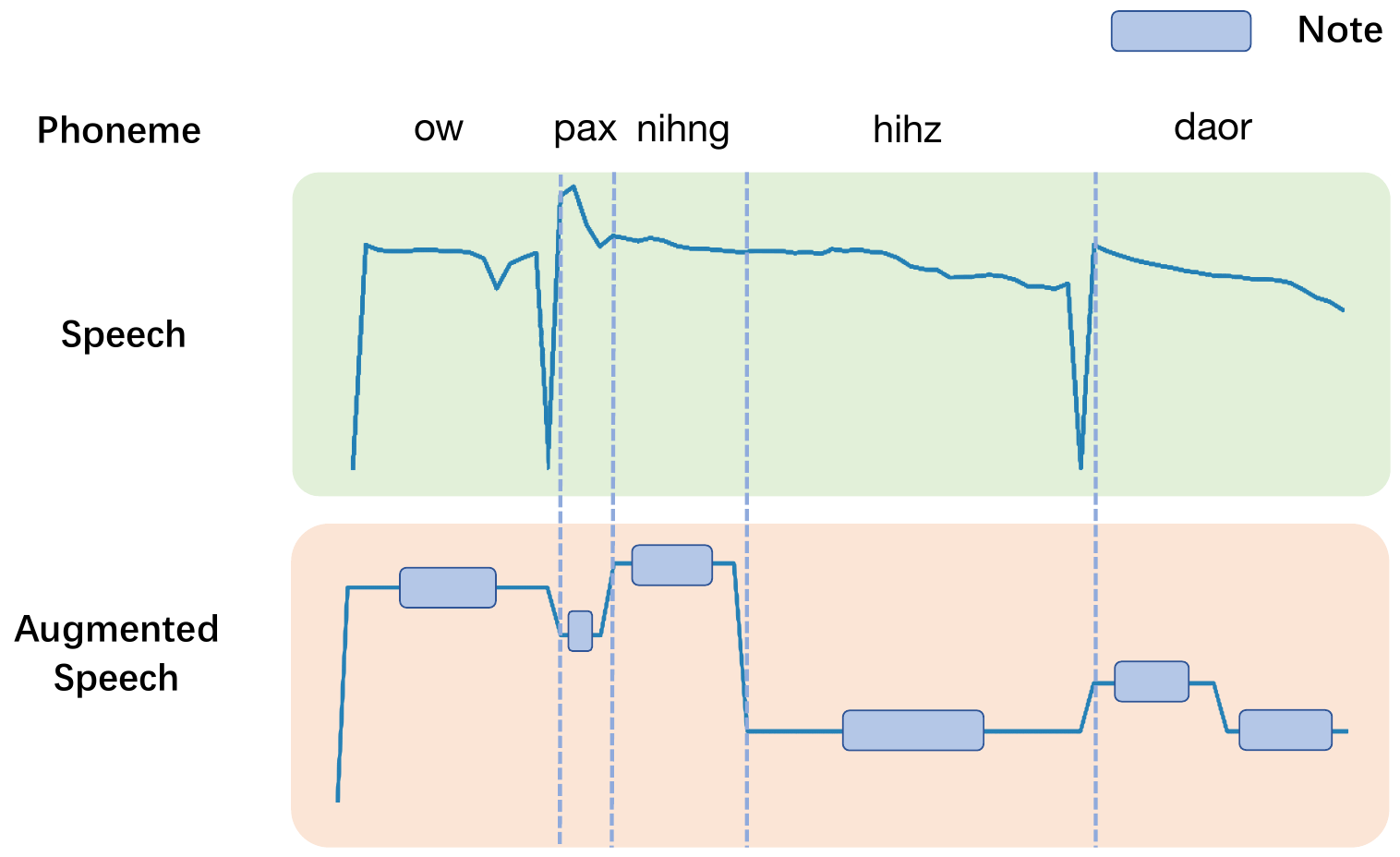}
    \caption{The change of F0 contour after pitch adjuster. The content of this example is “opening his door”.}
    \label{fig:pitch}
\end{figure}

Pitch adjuster adjusts the pitch of input speech at syllable level according to the aligned notes. Specifically, we use WORLD~\citep{morise2016world}, a fast and high-quality vocoder-based speech synthesis system to implement the adjustment. The WORLD system\footnote{https://github.com/JeremyCCHsu/Python-Wrapper-for-World-Vocoder} parameterizes speech into three components: fundamental frequency (F0), aperiodicity, and spectral envelope and can reconstruct the speech with only estimated parameters. We use WORLD to estimate the three parameters of natural speech and only adjust the F0 contours according to that of corresponding notes. Then we synthesize speech with adjusted F0 accompanied with original aperiodicity and spectral envelope.
Figure~\ref{fig:pitch} shows the F0 contours before and after pitch adjuster. 

Pitch adjuster calculates the pitch difference between speech and note with syllable-level alignment and adjusts F0 contours of speech accordingly. Some details are as follows:
1) Considering that the quality of synthesized speech will drop sharply when the range of adjustment is too large, we need to keep it within a reasonable threshold. Specifically, we calculate the average pitch of the speech and the corresponding melody respectively. When the average pitch of speech is too different from that of the corresponding melody (e.g., exceeding a threshold, which is 5 in our experiment), we shift the pitch of the entire note sequence to make the difference within the threshold and use the shifted note for adjustment, otherwise, keep the pitch of the original note unchanged;
2) To maintain smooth transitions in synthesized speech and prevent speech from being interrupted, we perform pitch interpolation for the frames between two syllables.
3) When a syllable is mapped to multiple notes, we segment the speech of this syllable in proportion to the duration of notes and adjust the pitch of each segment according to the corresponding note.

\subsection{Duration Adjuster}
\label{sec:aug_duration}
\begin{figure}[!thb]
    \centering
    \includegraphics[width=0.45\textwidth]{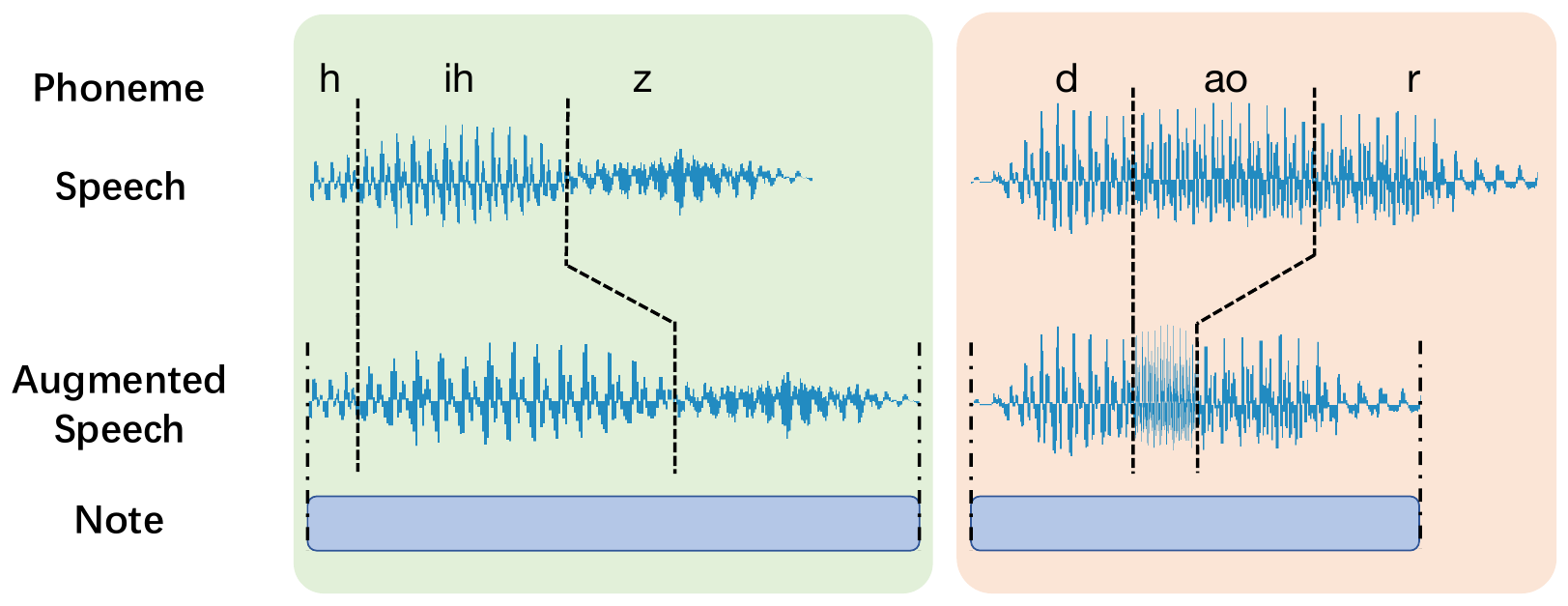}
    \caption{The change of duration after duration adjuster. The content of this example is “his door”. The left block shows the case of lengthening the duration of speech and the right shows the case of shortening the duration.}
    \label{fig:duration}
\end{figure}

Duration adjuster changes the duration of input speech to align with the duration of the corresponding note.
As shown in Figure~\ref{fig:duration}, instead of scaling the whole syllable, we only scale the length of vowels and keep the length of consonants unchanged, because the duration of consonants in singing voice is not significantly longer than that in speech, while long vowels are common in singing voice~\citep{kruspe2015training}. 
There are one-to-many mappings and many-to-one mappings in the syllable-level alignment. When multiple syllables are mapped to one note, we calculate the total length of the syllables and adjust the length of all vowels in these syllables in proportion. When multiple notes are mapped to one syllable, we adjust the length of the vowel, so that the total length of this syllable is equal to the total length of these notes.

\section{Experimental Settings}
In this section, we describe the experimental settings, including datasets, model configuration, and the details of training, inference, and evaluation.
\subsection{Datasets}
\paragraph{Singing Voice Datasets} We conduct experiments on two singing voice datasets to verify the effectiveness of our PDAugment: DSing30 dataset~\citep{dabike2019automatic} and Dali corpus~\citep{meseguer2019dali}. DSing30 dataset consists of about 4K monophonic Karaoke recordings of English pop songs with nearly 80K utterances, performed by 3,205 singers. We use the partition provided by~\citet{dabike2019automatic} to make a fair comparison with them. Dali corpus is another large dataset of synchronized audio, lyrics, and notes. It consists of 1200 English polyphonic songs with a total duration of 70 hours. Following~\citet{basak2021end}, we use the sentence-level annotation of the dataset provided by~\citet{meseguer2019dali}\footnote{https://github.com/gabolsgabs/DALI} and divide the dataset into training, development, and test with a proportion of 8:1:1 without any singers overlapping in each partition. We convert all the singing voice waveforms in our experiments into mel-spectrogram following~\citet{zhang2020uwspeech} with a frame size of 25 ms and hop size of 10 ms.
\paragraph{Natural Speech Dataset} Following the common practice in previous ASR work~\citep{amodei2016deep,gulati2020conformer,zhang2020transformer}, we choose the widely used LibriSpeech~\citep{panayotov2015librispeech} corpus as the natural speech dataset in our experiments. The LibriSpeech corpus contains 960 hours of speech sampled at 16 kHz with 1129 female speakers and 1210 male speakers. We use the official training partition described in~\citet{panayotov2015librispeech}. Similar to singing voice, we convert the speech into mel-spectrogram with the same setting.
\paragraph{Music Score Dataset} 
In this work, we choose the FreeMidi dataset\footnote{https://freemidi.org} and only use the pop songs among them\footnote{https://freemidi.org/genre-pop} because almost all of the songs in our singing voice datasets are pop music. The pop music subset of FreeMidi has about 4000 MIDI files, which are used to provide note information for PDAugment module. We use miditoolkit\footnote{https://github.com/YatingMusic/miditoolkit} to extract the note information from MIDI files and then feed them into the PDAugment module along with natural speech.
\paragraph{Lyrics Corpus}
\label{sec:data_lm}
In order to construct a language model with more in-domain knowledge, we deliberately collected a large amount of lyrics data to build our lyrics corpus for language model training. 
Besides the training text of DSing30 dataset and Dali corpus, we collect lyrics of English pop songs from the Web. We crawl about 46M lines of lyrics and obtain nearly 17M sentences after removing the duplication. We attach a subset of the collected lyrics corpus in the supplementary material.

\subsection{Model Configuration} 
\paragraph{ALT Model} We choose Conformer encoder~\citep{gulati2020conformer} and Transformer decoder~\citep{vaswani2017attention} as the basic model architecture in our experiments since the effectiveness of the structure has been proved in ASR. We stack $N=12$ layers of Conformer blocks in encoder and $N=6$ layers of Transformer blocks in decoder. The hidden size of both Conformer blocks and Transformer blocks are set to $512$, and the filter size of the feed-forward layer is set to $2048$. The number of the attention head is set to $8$ and the dropout rate is set to $0.1$.
\paragraph{Language Model} Our language model is based on Transformer encoder with $16$ layers, $8$ heads of attention, filter size of $2048$, and embedding unit of $128$. The language model is pre-trained separately and then integrated with ALT model in the decoding stage.

\subsection{Training Details}
During ALT training, after PDAugment module, we apply the SpecAugment~\citep{park2019specaugment} and speed perturbation. We use SpecAugment with frequency mask from $0$ to $30$ bins, time mask from $0$ to $40$ frames, time warp window of $5$. The speed perturbation factors are set to $0.9$ and $1.1$. We train the ALT model for 35 epochs on 2 GeForce RTX 3090 GPUs with a batch size of 175K frames.
Following~\citet{zhang2020denoising}, we use Adam optimizer~\citep{kingma2014adam} and set $\beta_1$, $\beta_2$, $\varepsilon$ to $0.9$, $0.98$ and $10^{-9}$ respectively. We apply label smoothing with $0.1$ weight when calculating the $\mathcal{L}_{dec}$ and set the $\lambda$ described in Section~\ref{eq:loss} to $0.3$.
We train the language model for 25 epochs on 2 GeForce RTX 3090 GPUs with Adam optimizer.

Our code of basic model architecture is implemented based on the ESPnet toolkit~\citep{watanabe2018espnet}\footnote{https://github.com/espnet/espnet}. We attach our code of PDAugment module to the supplementary materials. 

\subsection{Inference and Evaluation}
During inference, we fuse the ALT model with the language model which is pre-trained with lyrics corpus. Following previous ALT work~\citep{demirel2020automatic,basak2021end}, we use the word error rate (WER) as the metric when evaluating the accuracy of lyrics transcription.

\section{Results and Analyses}
In this section, we first report the main experiment results, and then conduct ablation studies to verify the effectiveness of each component in PDAugment, finally analyze the adjusted speech by statistics and visualization.

\subsection{Main Results}
In this subsection, we report the experimental results of the ALT system equipped with PDAugment in two singing voice datasets. We compare our results with several basic settings as baselines: 1) \textit{Naive ALT}, the ALT model trained with only singing voice dataset; 2) \textit{ASR Augmented}, the ALT model trained with the combination of singing voice dataset and ASR data directly; 3) \textit{Random Aug}~\citep{kruspe2015training}, the ALT model trained with the combination of singing voice dataset and randomly adjusted ASR data. The pitch is adjusted ranging from -6 to 6 semitones randomly and the duration ratio of speech before and after adjustment is randomly selected from 0.5 to 1.2. All of 1), 2), and 3) are using the same model architecture as \textit{PDAugment}. Besides the above three baselines, we compare our PDAugment with the previous systems which reported the best results in two datasets respectively. For DSing30 dataset, we compare our PDAugment with~\citet{demirel2020automatic} using RNNLM\footnote{https://github.com/emirdemirel/ALTA} and the results are shown in Table~\ref{tab:dsing}. For Dali corpus, we compare the results with~\citet{basak2021end} and report the results in Table~\ref{tab:dali}.
\begin{table}[!h]
\centering
\small 
\begin{tabular}{l | c c }
\toprule
Method & \textit{Dev} & \textit{Test} \\
\midrule
\textit{Naive ALT} & 28.2 & 27.4 \\
\textit{ASR Augmented} & 20.8 & 20.5 \\
\textit{Random Aug~\citep{kruspe2015training}} & 17.9 & 17.6 \\
\textit{\citet{demirel2020automatic}} & 17.7 & 15.7 \\
\midrule
\textit{PDAugment} & \textbf{10.1} & \textbf{9.8} \\
\bottomrule
\end{tabular}
\caption{The WERs (\%) of DSing30 dataset.}
\label{tab:dsing}
\end{table}

\begin{table}[!h]
\centering
\small 
\begin{tabular}{l | c c }
\toprule
Method & \textit{Dev} & \textit{Test} \\
\midrule
\textit{Naive ALT} & 80.9 & 86.3 \\
\textit{ASR Augmented} & 75.5 & 75.7 \\
\textit{Random Aug~\citep{kruspe2015training}} & 69.8 & 72.1 \\
\textit{\citet{basak2021end}} & 75.2 & 78.9 \\
\midrule
\textit{PDAugment} & \textbf{53.4} & \textbf{54.0} \\
\bottomrule
\end{tabular}
\caption{The WERs (\%) of Dali corpus.}
\label{tab:dali}
\end{table}

As can be seen, \textit{Naive ALT} performs not well and gets high WERs in both DSing30 dataset and Dali corpus, which demonstrates the difficulty of the ALT task. After adding ASR data for ALT training, the performances of \textit{ASR Augmented} setting in both of the two datasets have been improved slightly compared to \textit{Naive ALT}, but still with relatively high WERs, which indicates the limitation of using ASR training data directly. 

When applying the adjustments, a question is that if we adjust the pitch and duration with random ranges without note information from music scores, how well will the ALT system perform? The results of \textit{Random Aug} can perfectly answer this question. As the results show, \textit{Random Aug} can slightly improve the performance compared with \textit{ASR Augmented}, demonstrating that increasing the volatility of pitch and duration in natural speech helps ALT system training, which is the same as what \citet{kruspe2015training} claimed. \textit{PDAugment} is significantly better than \textit{Random Aug}, which indicates that adjusted speech can better help the ALT training with the guidance of music scores.

Besides, it is obvious that \textit{PDAugment} greatly outperforms the previous SOTA in both datasets. \citet{demirel2020automatic} in Table~\ref{tab:dsing} performs worse than \textit{PDAugment} because of not taking advantage of the massive ASR training data. Compared with \citet{basak2021end} in Table~\ref{tab:dali} that replaced F0 contours of speech directly, \textit{PDAugment} can narrow the gap between natural speech and singing voice in a more reasonable manner and achieve the lowest WERs among all the above methods. The results in both datasets show the effectiveness of PDAugment for ALT task and reflect the superiority of adding music-specific acoustic characteristics into natural speech.

\begin{figure*}[!thb]
	\centering
	\begin{subfigure}[h]{0.24\textwidth}
	    \captionsetup{justification=centering}
		\centering
		\includegraphics[width=\textwidth, height=0.7\textwidth, trim={0.5cm 0cm 2cm 1.8cm}, clip=true]{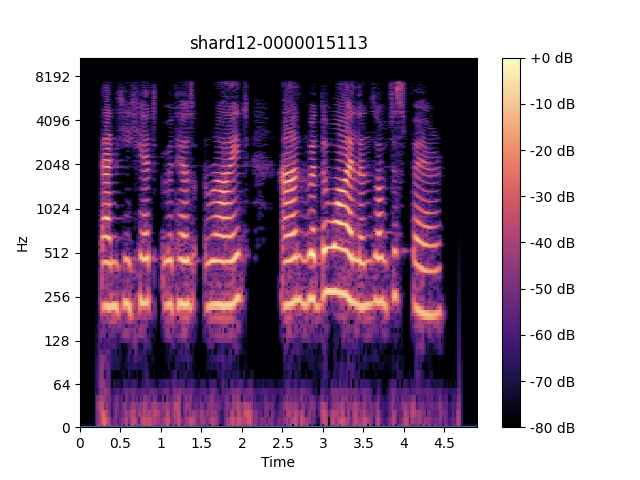}
		\caption{Original speech sample.}
		\label{fig:ori_speech}
	\end{subfigure}
	\begin{subfigure}[h]{0.24\textwidth}
	    \captionsetup{justification=centering}
		\centering
		\includegraphics[width=\textwidth, height=0.7\textwidth, trim={0.5cm 0cm 2cm 1.8cm}, clip=true]{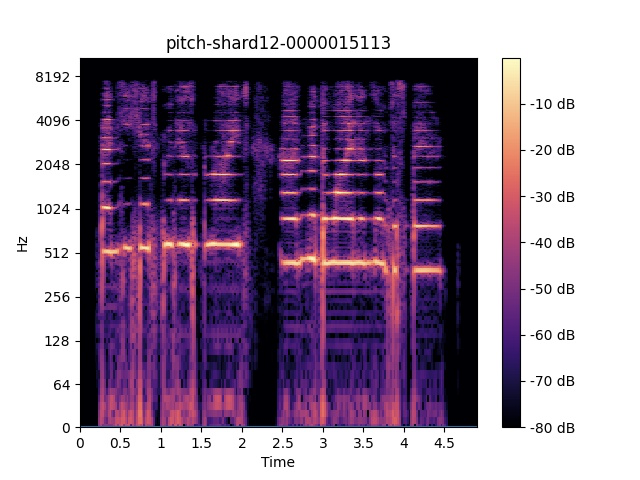}
		\caption{After Pitch Adjuster.}
		\label{fig:aug_pitch}
	\end{subfigure}
	\begin{subfigure}[h]{0.24\textwidth}
	    \captionsetup{justification=centering}
		\centering
		\includegraphics[width=\textwidth, height=0.7\textwidth, trim={0.5cm 0cm 2cm 1.8cm}, clip=true]{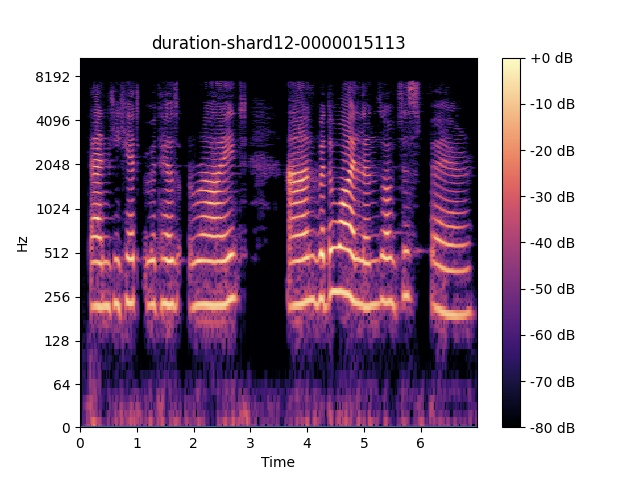}
		\caption{After Duration Adjuster.}
		\label{fig:aug_duration}
	\end{subfigure}
	\begin{subfigure}[h]{0.24\textwidth}
	    \captionsetup{justification=centering}
		\centering
		\includegraphics[width=\textwidth, height=0.7\textwidth, trim={0.5cm 0cm 2cm 1.8cm}, clip=true]{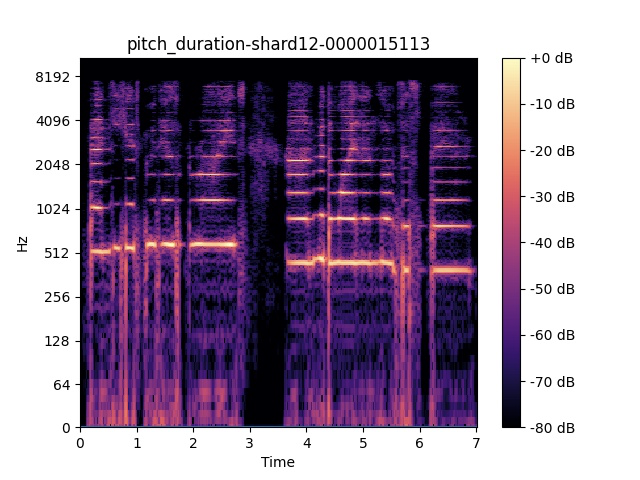}
		\caption{After PDAugment.}
		\label{fig:aug_pd}
	\end{subfigure}
	\caption{Spectrograms of speech example after pitch adjuster or/and duration adjuster.}
	\label{fig:aug_example}
\end{figure*}

\subsection{Ablation Studies}
We conduct more experimental analyses to deeply explore our PDAugment and verify the necessity of some design details. More ablation studies (different language models) can be found in the supplementary materials. The ablation studies are carried out on DSing30 dataset in this section.

\subsubsection{Augmentation Types} 
In this subsection, we explore the effects of different kinds of augmentations (only adjust pitch, only adjust duration, and adjust pitch \& duration) on increasing the performance of ALT system.
We generate three types of adjusted speech by enabling different adjusters of PDAugment module and conduct the experiments on DSing30. The results are shown in Table~\ref{tab:diff_aug}.

\begin{table}[!h]
\centering
\small 
\begin{tabular}{ l | c c }
\toprule
Setting & \textit{DSing30 Dev} & \textit{DSing30 Test} \\
\midrule
\textit{PDAugment} & \textbf{10.1} & \textbf{9.8} \\
\textit{- Pitch Adjuster} & 13.6 & 13.4 \\
\textit{- Duration Adjuster} & 13.8 & 13.8 \\
\textit{- Pitch \& Duration Adjusters} & 20.8 & 20.5 \\
\bottomrule
\end{tabular}
\caption{The WERs (\%) of different types of augmentation of DSing30 dataset. All of the settings are trained on DSing30 and the original or adjusted LibriSpeech data.}
\label{tab:diff_aug}
\end{table}

As we can see, when we enable the whole PDAumgnet module, the ALT system can achieve the best performance, indicating that the effectiveness of the pitch and duration adjusters. When we disable the pitch adjuster, the WER on DSing30 is 3.6\% higher than \textit{PDAugment}. The same thing happens when we disable the duration adjuster, the WER is 4.0\% higher than \textit{PDAugment}. And if both the pitch and duration are not adjusted, which means use the speech data directly for ALT training, the WER is the worst among all settings. The results demonstrate that both pitch and duration adjusters are necessary and can help with improving the recognition accuracy of the ALT system.

\subsection{Adjusted Speech Analyses}
\subsubsection{Statistics of adjusted Speech}
Following Section~\ref{sec:speech_vs_singingvoice}, we analyze the acoustic properties of the original natural speech and the adjusted speech by PDAugment, and list the results in Table~\ref{tab:statistics2}. 

\begin{table}[!h]
\centering
\small 
\begin{tabular}{l | c c }
\toprule
Property & \textit{Original Speech} & \textit{adjusted Speech} \\
\midrule
\textit{Pitch Range (semitone)} & 12.71 & 14.19 \\
\textit{Pitch Smoothness} & 0.93 & 0.69 \\
\textit{Duration Range (s)} & 0.44 & 0.59 \\
\textit{Duration Variance} & 0.01 & 0.05 \\
\bottomrule
\end{tabular}
\caption{The differences of acoustic properties between original speech and adjusted speech.}
\label{tab:statistics2}
\end{table}

Combining the information of Table~\ref{tab:statistics} and Table~\ref{tab:statistics2}, we can clearly find that the distribution pattern of acoustic properties (pitch and duration) after PDAugment is closer to singing voice compared with the original speech, which indicates that our PDAugment can change the patterns of pitch and duration in original speech and effectively narrow the gap between natural speech and singing voice. To avoid the distortion of adjusted speech, we limit the adjustment degree within a reasonable range, so the statistics of adjusted speech can not completely match that of singing voice. Nonetheless, adjusted speech is still good enough for ALT model to capture some music-specific characteristics.

\subsubsection{Visualization of adjusted Speech}
In order to visually demonstrate the effect of PDAugment module, we plot the spectrograms of speech to compare the acoustic characteristics before and after different types of adjustments.

An example of our PDAugment output is shown in Figure~\ref{fig:aug_example}. In detail, the spectrogram of the original natural speech sample is shown in Figure~\ref{fig:ori_speech}. And we illustrate the spectrograms of purely pitch adjusted, purely duration adjusted, both pitch and duration adjusted speech in Figure~\ref{fig:aug_pitch}, Figure~\ref{fig:aug_duration}, and Figure~\ref{fig:aug_pd} separately. 
It is clear that the adjusted speech audios have different acoustic properties from natural speech audios. 
More example spectrograms and audios can be found in the supplementary material.

\section{Conclusion}
In this paper, we proposed PDAugment, a data augmentation method by adjusting pitch and duration to make better use of natural speech for ALT training. PDAugment transfers natural speech into singing voice domain by adjusting pitch and duration at syllable level under the instruction of music scores. PDAugment module consists of speech-note aligner to align the speech with note, and two adjusters to adjust pitch and duration respectively. Experiments on two singing voice datasets show that PDAugment can significantly reduce the WERs of ALT task. Our method analyses explore different types of augmentation, further verify the effectiveness of PDAugment. In the future, we will consider narrowing the gap between natural speech and singing voice from more aspects such as vibrato and try to add some music-specific constraints in the decoding stage.

\bibliography{aaai22}

\end{document}


\maketitle

\section{Ablation Studies on Language Models}
There are some discrepancies between the lyrics domain and the natural text domain, so the language models used for ALT and ASR should be different. In this part, we explore how different language models will influence the performance of ALT system. We compare the language model trained on LibriSpeech~\citep{panayotov2015librispeech} data with that trained on our collected lyrics corpus. 
As shown in Table~\ref{tab:diff_lm}, compared to \textit{Without LM}, both language models can improve the transcription accuracy. \textit{LM on lyrics corpus} outperforms \textit{LM on LibriSpeech}, which indicates that the language model trained on in-domain lyrics data indeed better modeling linguistic information for ALT.
\begin{table}[!h]
\centering
\small 
\begin{tabular}{l | c c }
\toprule
Setting & \textit{DSing30 Dev} & \textit{DSing30 Test} \\
\midrule
\textit{Without LM} & 12.5 & 12.4 \\
\textit{LM on LibriSpeech} & 12.5 & 12.2 \\
\textit{LM on lyrics corpus} & \textbf{10.1} & \textbf{9.8} \\
\bottomrule
\end{tabular}
\caption{The WERs (\%) of different language models.}
\label{tab:diff_lm}
\end{table}

\section{Adjusted Speech Analyses}
Besides the spectrograms we illustrate in Section 5.3, we save spectrograms of more examples in MultimediaAppendix \textbf{\textit{spectrograms}} and audios of those examples in MultimediaAppendix \textbf{\textit{audios}}. Both of the \textbf{\textit{spectrograms}} and \textbf{\textit{audios}} directories have four subfolders: 1) \textbf{\textit{Original Speech}}, which contains the spectrograms or audios of the original speech; 2) \textbf{\textit{Duration Adjuster}}, which contains the spectrograms or audios of the adjusted speech after duration adjuster; 3) \textbf{\textit{Pitch Adjuster}}, which contains the spectrograms or audios of the adjusted speech after pitch adjuster; and 4) \textbf{\textit{PDAugment}}, which contains the spectrograms or audios of the adjusted speech after the entire PDAugment module.

\section{Data and Codes}
\subsection{Lyrics Corpus}
We crawl about 46M lines of lyrics from the Web and obtain nearly 17M sentences after removing the duplication. 
We randomly select 2000 sentences from the whole lyrics corpus
and save subset of the lyrics corpus we collected in CodeAndDataAppendix \textbf{\textit{lyrics\_corpus.txt}}.
\subsection{Code}
Our code of basic model architecture is implemented based on the ESPnet toolkit (Watanabe et al. 2018). We attach our code of PDAugment module to CodeAndDataAppendix \textbf{\textit{codes}}, and the details of our code are explained in CodeAndDataAppendix \textbf{\textit{codes/README.md}}

\bibliography{aaai22}